\documentclass[american,aps,prl,reprint,showpacs,superscriptaddress,altaffilletter]{revtex4-1}
\usepackage{lmodern}
\usepackage{lmodern}
\usepackage[T1]{fontenc}
\usepackage[utf8]{inputenc}
\usepackage{xcolor}
\usepackage{babel}
\usepackage{amsmath}
\usepackage{amssymb}
\usepackage{graphicx}
\usepackage{wasysym}
\usepackage[unicode=true,
 bookmarks=true,bookmarksnumbered=false,bookmarksopen=false,
 breaklinks=false,pdfborder={0 0 0},pdfborderstyle={},backref=false,colorlinks=true]
 {hyperref}
\hypersetup{
 pdfborderstyle={},pdfborderstyle={},pdfborderstyle={},pdfborderstyle={},pdfborderstyle={},pdfborderstyle={},linkcolor=blue,citecolor=blue}

\makeatletter
\usepackage{comment}

\makeatother

\begin{document}
%\title{Tuning the restitution coefficient  of closed containers partially filled with water}
\title{Fluid motion for reducing the bounce of partially filled containers}

\author{Klebbert Andrade}
\affiliation{Departamento de Física, Facultad de Ciencia, Universidad de Santiago de Chile, ~\\
 Av. Víctor Jara 3493, Estación Central, Santiago, Chile}

\author{Javiera Catal\'an}
\affiliation{Departamento de Física, Facultad de Ciencia, Universidad de Santiago de Chile, ~\\
 Av. Víctor Jara 3493, Estación Central, Santiago, Chile}

\author{Juan F. Mar\'in}
\affiliation{Departamento de Física, Facultad de Ciencia, Universidad de Santiago de Chile, ~\\
 Av. Víctor Jara 3493, Estación Central, Santiago, Chile}
\affiliation{Laboratoire Gulliver, ESPCI-Paris, Université PSL, 10 rue Vauquelin, 75005 Paris, France}

\author{Vicente Salinas}
%\email{vicente.salinas@uautonoma.cl}
\affiliation{Grupo de Investigaci\'on en F\'isica Aplicada, Instituto de Ciencias Qu\'imicas Aplicadas, Facultad de Ingenier\'ia, Universidad Aut\'onoma de Chile, Santiago, Chile}

\author{Gustavo Castillo}
%\email{gustavo.castillo@uoh.cl}
\affiliation{Instituto de Ciencias de la Ingeniería, Universidad de O'Higgins,
Av. Libertador Bernardo O'Higgins 611, Rancagua, Chile}

%\author{Francisco Olea}
%\affiliation{Departamento de Física, Universidad de Santiago de Chile, ~\\
%Av. Ecuador 3493, Estación Central, Santiago, Chile}

%\author{Enriko Granadoz}
%\affiliation{Departamento de Física, Universidad de Santiago de Chile, ~\\
%Av. Ecuador 3493, Estación Central, Santiago, Chile}

\author{Leonardo Gordillo}
\email{leonardo.gordillo@usach.cl}
\affiliation{Departamento de Física, Facultad de Ciencia, Universidad de Santiago de Chile, ~\\
 Av. Víctor Jara 3493, Estación Central, Santiago, Chile}
 
 \author{Pablo Guti\'errez}
\email{pablo.gutierrez@uoh.cl}
\affiliation{Instituto de Ciencias de la Ingeniería, Universidad de O'Higgins,
Av. Libertador Bernardo O'Higgins 611, Rancagua, Chile}

\begin{abstract} %OJO que ahora PRL pide abstract < 600 CARACTERES
Certain spatial distributions of water inside partially filled containers can significantly reduce the bounce of the container. In experiments with containers filled to a volume fraction $\phi$, we show that rotation offers control and high efficiency in setting such distributions and, consequently, in altering bounce markedly. High-speed imaging evidences the physics of the phenomenon and reveals a rich sequence of fluid-dynamics processes, which we translate into a model that captures our overall experimental findings.
\end{abstract}

%We demonstrate that the bounce of containers partially filled with water can be significantly reduced by properly distributing the water inside the container before impact. We show experimentally that simply rotating the container prior to release offers  straightforward control of the initial distribution of water as a function of the rotation frequency $\omega$ and the filling volume fraction $\phi$ of the container. We accordingly characterize the bounce of containers set into rotation via their restitution coefficient $e$, finding optimal values for $\phi$ and $\omega$ where maximal bounce reduction is attained.
%High-speed imaging evidences the physics that reduce the bounce when rotation is present and reveals a rich sequence of fluid-dynamics processes, which we translate into a model that overall captures our experimental findings.

%PACS: 89.75.Kd: Pattern formation in complex systems; 47.55.np Contact
%lines; 47.35.Bb Gravity waves; 47.15.ki Inviscid flows with vorticity;
%47.20.Ft Instability of shear flows

\maketitle

%\paragraph{Introduction\label{sec:intro}}

{The impact of an elastic container partially filled with liquid radically differs from the impact of both elastic solids and unconstrained liquids. During impact, elastic solids deform, but their shape is recovered almost unaffected after detachment \cite{Stronge}. As a consequence, elastic solids display big bounces and almost perfect elastic collisions. Unconstrained liquids otherwise undergo large and irreversible deformations during impact. Liquids rarely bounce \cite{Rein_1993,Biance_2006,Zhang_2022}, instead, they spread on the impact zone and form lamellae, fingers, or jets \cite{Rein_1993,Yarin_2006,JosserandThoroddsen_2016}. As this happens, they exert signature forces on the surface \cite{GordilloEtAl_2018,ARFMCheng_2022,NatureComm_Sun_2022,Zhang_2022}, in a process that could be labeled as perfectly inelastic.  Containers partially filled with a liquid behave uniquely. Although there is available space for spreading, redistribution remains bounded by the walls allowing momentum transfer between the liquid and the container. The whole system dynamics and outcome after impact are no longer easy to predict \cite{PachecoVazquezEtAl_2014}.}

{An early study on the impact of an open, cylindrical container partially filled with liquid \cite{Milgram_1969}, revealed the ubiquity of a focused central liquid jet after impact.  The jet intensity has been shown to depend on the shape of the liquid surface just before the impact \cite{AntkowiakEtAl_2007}. The jet naturally carries away a part of the momentum and energy of the container after contact. This interaction creates strange bouncing patterns of the whole system, as those experienced by partially filled spheres \cite{KillianEtAl_2012}. When studying the whole system dynamics, it is useful to introduce the restitution coefficient $e = |\, v_+ /v_- |$, where $v_-$ is the container velocity prior to impact, and $v_+$, the one after \cite{Stronge, FalconEtAl_1998, PachecoVazquez_2013, GarciaCidEtAl_2015}. For the sphere \cite{KillianEtAl_2012}, they found that $e$ strongly depends both on the state of the liquid surface (whether it was perturbed out of equilibrium or not, prior to impact) and on the sphere filling volume fraction. On the other hand, $e$ barely depends on the physical properties of the fluid or the container. The mechanism contrasts with those observed in containers partially filled with grains, where bounce can be fully attenuated if the container is properly filled, due to the highly dissipative nature of collisions  \cite{PachecoVazquez_2013}. It should be noted that no control of the bouncing containers has ever been documented for containers filled with liquid.}

\begin{figure}[b]%[htbp]
\begin{centering}
\includegraphics[width=1\columnwidth]{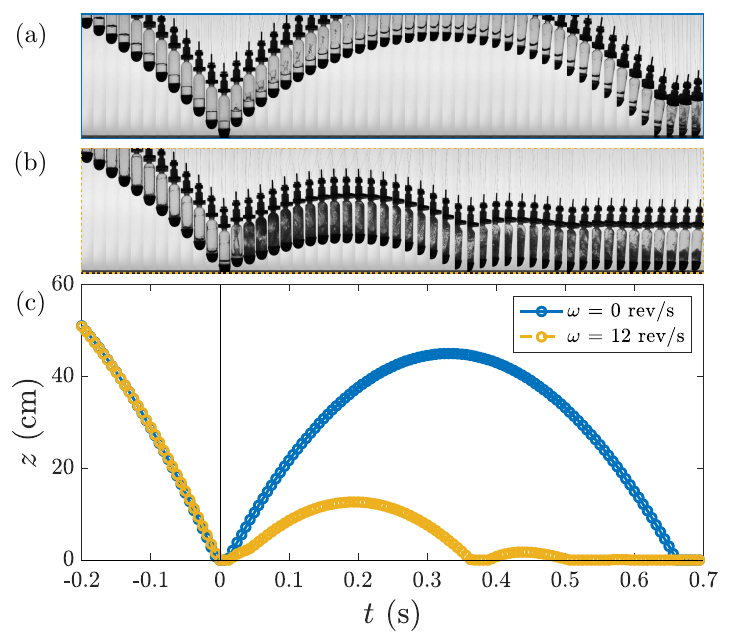}
\par\end{centering}
\caption{Impact of partially filled containers. (a) Sequence of images for a container dropped with a quiescent fluid inside. (b) Sequence when the container is rotated at frequency $\omega$~=~12~rev/s before release. (c) Trajectories for the lowest point of the container, for the same time interval as panels (a) and (b).\label{Fig:SummaryAndSetup}}
\end{figure}

{This letter aims to present a simple fluid--mechanical way to tune the restitution coefficient for partially filled containers employing a preset motion. The combination of available volume within the container, and an imposed rotation before the release, creates fluid distributions prior to the impact that tune the bounce in exceptional ways. Figure \ref{Fig:SummaryAndSetup} summarizes our findings by comparing a container dropped with a quiescent fluid inside (Fig.~\ref{Fig:SummaryAndSetup}a) with another one dropped after a strong rotation was set up %as visible in the parabolic shape of the free surface before release 
(Fig.~\ref{Fig:SummaryAndSetup}b). The bounce in the rotating case is largely attenuated, compared with the quiescent fluid counterpart, as shown by the $z$ vs. $t$ curves in Fig.~\ref{Fig:SummaryAndSetup}c. Surprisingly, increasing the total energy injected into the system via rotational kinetic energy reduces the container bounce.}

\begin{figure}[b]
\begin{centering}
\includegraphics[width=0.85\columnwidth]{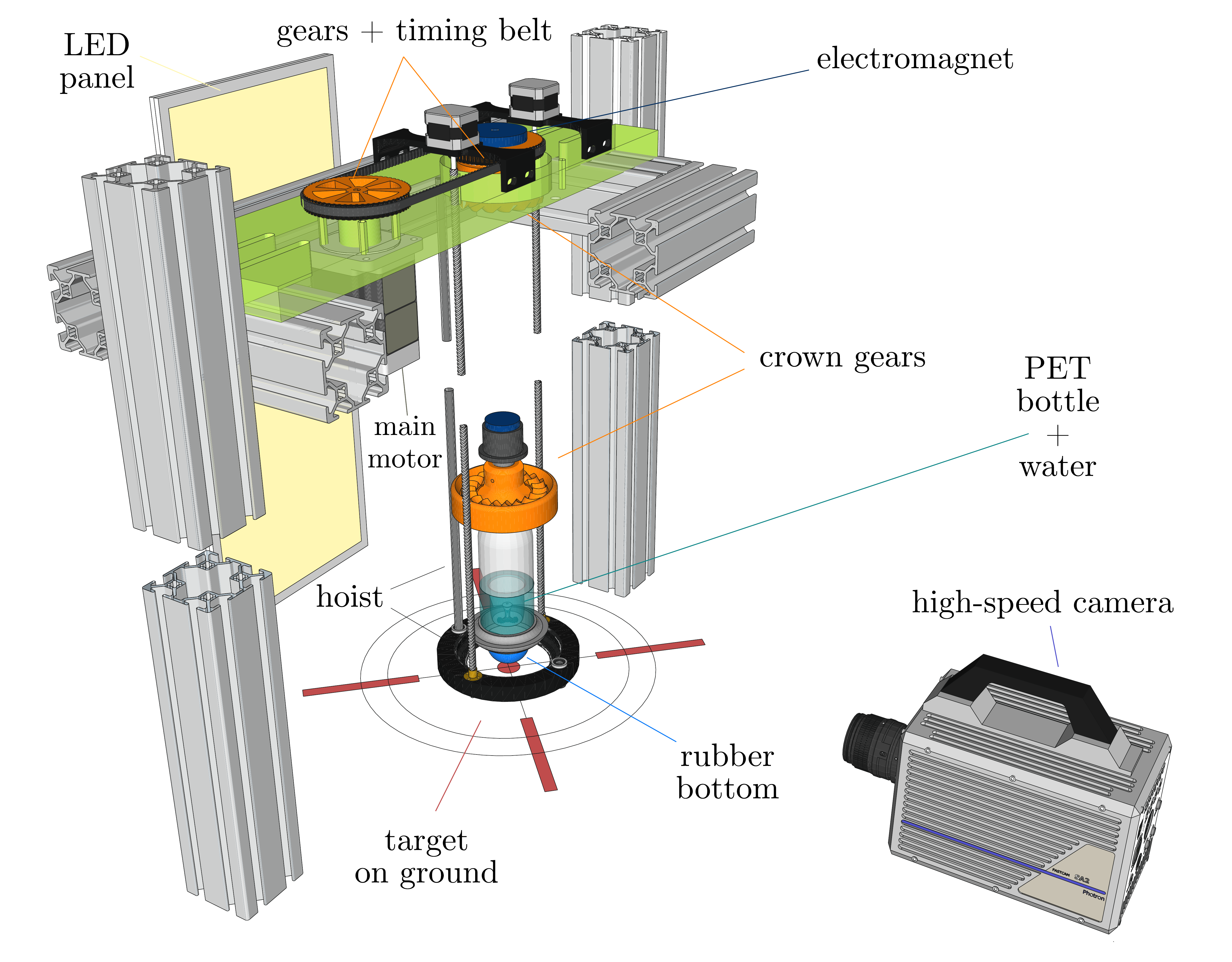}
\par\end{centering}
\caption{The experimental setup consists of a system for bottle rotation and release, video sequencing of impact, and a hoist for relocating the bottle and resuming measurements. The discontinuity indicates that the release height is much larger than shown. \label{Fig:Setup}}
\end{figure}
%Figure \ref{Fig:SummaryAndSetup}

\paragraph{}
{To quantify the restitution coefficient $e$, we performed experiments with a container partially filled with water that impacts a target after falling a height $H$. We used two control parameters: the containers’ initial angular velocity $\omega$ and its filling volume fraction $\phi$. The initial angular velocity provides an easy and effective way to control the water distribution before the impact. As we present in Fig. \ref{Fig:Setup}, container rotation is achieved via a stepper motor (Parker HV223), which transmits motion to the container axis through an arrangement of gears and belts. We established a protocol of a gradual rotation increase followed by an abrupt stop, after which we released the container using an electromagnet and a relay. While falling, the container activated a photogate that triggers the acquisition of images. We used a Phantom 410S high-speed camera, running at 2000 fps.
After an impact event, a tailored hoist lifts the container until it reaches the electromagnet again, resetting a new cycle of the experiment. The whole process (both control and acquisition) runs autonomously using a Matlab\textregistered~code. From images as those presented in Fig.~\ref{Fig:SummaryAndSetup}, we determine the velocity of the container before ($v_-$) and after impact ($v_+$), obtaining the restitution coefficient $e$.}

{The container is a cylindrical PET bottle (60-mm \diameter~and $h=$18-cm height), with an elastic half-sphere ({57-mm~\diameter}) glued to its bottom with the purpose of maximizing bounce, and thus highlighting the influence of the liquid. The bottle is sealed with a specially devised magnetic cap that also sticks to the electromagnet and allows mechanical coupling with the rotation system through a bearing and a crown gear (see Fig. \ref{Fig:Setup}). The bottle has a total mass of $m_b=584$ gr, an available volume of $V_b=538$ cm$^3$ and can be filled with tap water with masses going from zero to 528 gr ($\phi$ = 1). The bottle is released a fixed height of $H=$ 73.4 cm, reaching impact velocities around 340 cm/s when hitting the target.}

\begin{figure}[tb]
\begin{centering}
\includegraphics[width=0.95\columnwidth]{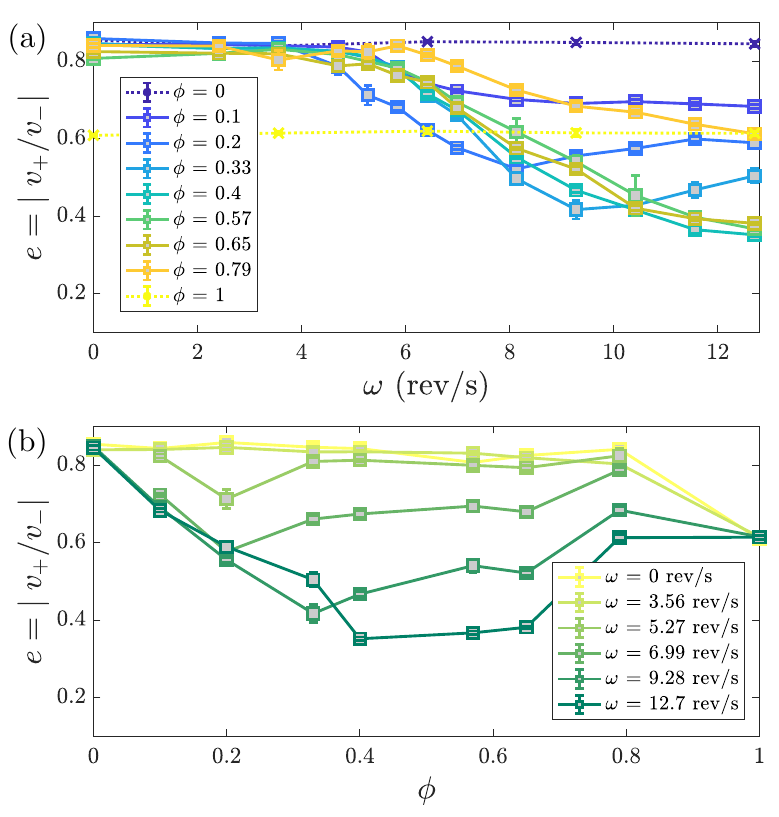}
\par\end{centering}
\caption{Restitution coefficient quantifying the water-damping effect. (a) Dependence of the restitution coefficient as a function of the bottle rotation $\omega$ before release. (b) Dependence of the restitution coefficient as a function of the bottle filling volume fraction $\phi$.}\label{Fig:CORvsOmega}
\end{figure}

{To quantify the water-damping effect on the container in terms of $\omega$ and $\phi$, we use the restitution coefficient $e$. Figure \ref{Fig:CORvsOmega}a shows its dependence on $\omega$ for various fixed filling volume fractions $\phi$, represented by different colors. Each data point represents the mean value of at least five realizations, while error bars were obtained from standard deviations. Two control cases are presented as dotted lines: an empty ($\phi$~=~0) and brimful bottle ($\phi$~=~1), for which no significant motion inside the container is possible. As expected, $\omega$ barely affects the motion in these cases. For $\phi$ = 1, plastic deformations on the container walls result in a different source of bounce attenuation and a change of trend.
More interestingly, for intermediate filling volume fraction, %($\phi \approx 0.5$), esto es valido en todo el rango, me parece innecesariamente confuso restringirlo a 0.5.
$e$ substantially drops as $\omega$ increases. Impacts with large liquid volumes ($\phi\gtrsim 0.8$) also display a monotonic reduction of their restitution coefficient $e$ but to a lesser degree. For lower filling volume fraction $\phi$ (up to 0.33), the measurements show some increment of $e$ above a minimal optimal value $\omega_{min}$. The trends of the set of measurements suggest that all the curves reach a plateau at some large $\omega$. 
 %The plateau values converge to the restitution coefficient of the brimful bottle as $\phi$ approaches 1. For low filling fraction $\phi$ (up to 0.33), the series of measurements shows an optimal $e$ for which the restitution coefficient is minimal. 
 Experimental data thus supports that controlling $\omega$ can tune the overall restitution coefficient of the system.}

{The results moreover indicate that the control range of $\omega$ on the restitution coefficient $e$ depends strongly on the filling volume fraction $\phi$. This is further evidenced when plotting $e$ as a function of the filling fraction at fixed $\omega$, as shown in Fig.~\ref{Fig:CORvsOmega}b. A minimum $e$ at some given $\phi$ is always observed, and in most cases at values between 0.2 and 0.4, which means that the optimal water-damping requires a significant proportion of available volume inside the container.}
{These results are close to those of Killian, Klaus \& Truscott, who suggested an optimal filling volume fraction of approximately 30\% for partially filled spheres under successive rebounce \cite{KillianEtAl_2012}.}

\paragraph{}
%Figure \ref{Fig:SummaryAndSetup} already showed the main effect we are interested in: the attenuation of the container’s bounce when water rotates before the impact. However, we can further quantify the effect summarized by the restitution coefficient $e$ while varying the control parameters $\omega$ and $\phi$, which we present in Fig. 

%{Figure \ref{Fig:SummaryAndSetup} already shows the main effect we are interested in: the damping of the container’s bounce when water rotates before the impact. %
{To deepen into the physical mechanism of the bounce reduction, we analyzed the high-speed video sequences \cite{SuppMat}. We start by comparing the fluid motion of the experiments of Fig.~\ref{Fig:SummaryAndSetup}. When there is no rotation (Fig.~\ref{Fig:SummaryAndSetup}a), water keeps at rest while freely falling, and a slow and weak bump on the surface forms during the impact. The bump develops into a low-speed jet only later. The course is similar to those in impacts with filled open containers \cite{Milgram_1969,AntkowiakEtAl_2007}. In contrast, water dynamics is anomalously richer when the container rotates before the release (Fig.~\ref{Fig:SummaryAndSetup}b). We describe the main stages of this process in a comoving frame of reference, as depicted in Fig.~\ref{Fig:StagesScheme}. (\emph{i}) Before the release ($t=t_r^-$), the water inside the bottle is at a steady rotational state, and its free surface has the expected shape of a paraboloid. (\emph{ii}) Right after the release ($t=t_r^-$), water stops experiencing gravity and starts to climb up on the container walls solely propelled by the action of the centrifugal force (see also Fig.~\ref{Fig:StagesScheme}b). The waterfront spreads upward with a velocity that depends on the rotation. Therefore, how much water covers the walls right before the impact ($t=0^-$) depends on how fast it rotates, the bottle fall time, and the filling volume fraction. (\emph{iii}) During the contact ($t=0^+$), while the elastic sphere is compressed, the water descends rapidly on the walls. A central jet emerges nourished by the incoming water from the walls and focused on the axis, and as a result, an effective force is exerted downward on the bottom of the bottle (see also Fig.~\ref{Fig:StagesScheme}c). Thus, the bounce after the impact depends dramatically on how much water is available to descend, and hence $e$ depends on the initial angular velocity. Also, the jet is much faster than in the $\omega=0$ case, and much thicker because of the large centrifugal force due to the angular momentum transferred from the water rotating originally on the wall. 
\begin{figure}[t!]
\begin{centering}
\includegraphics[width=1\columnwidth]{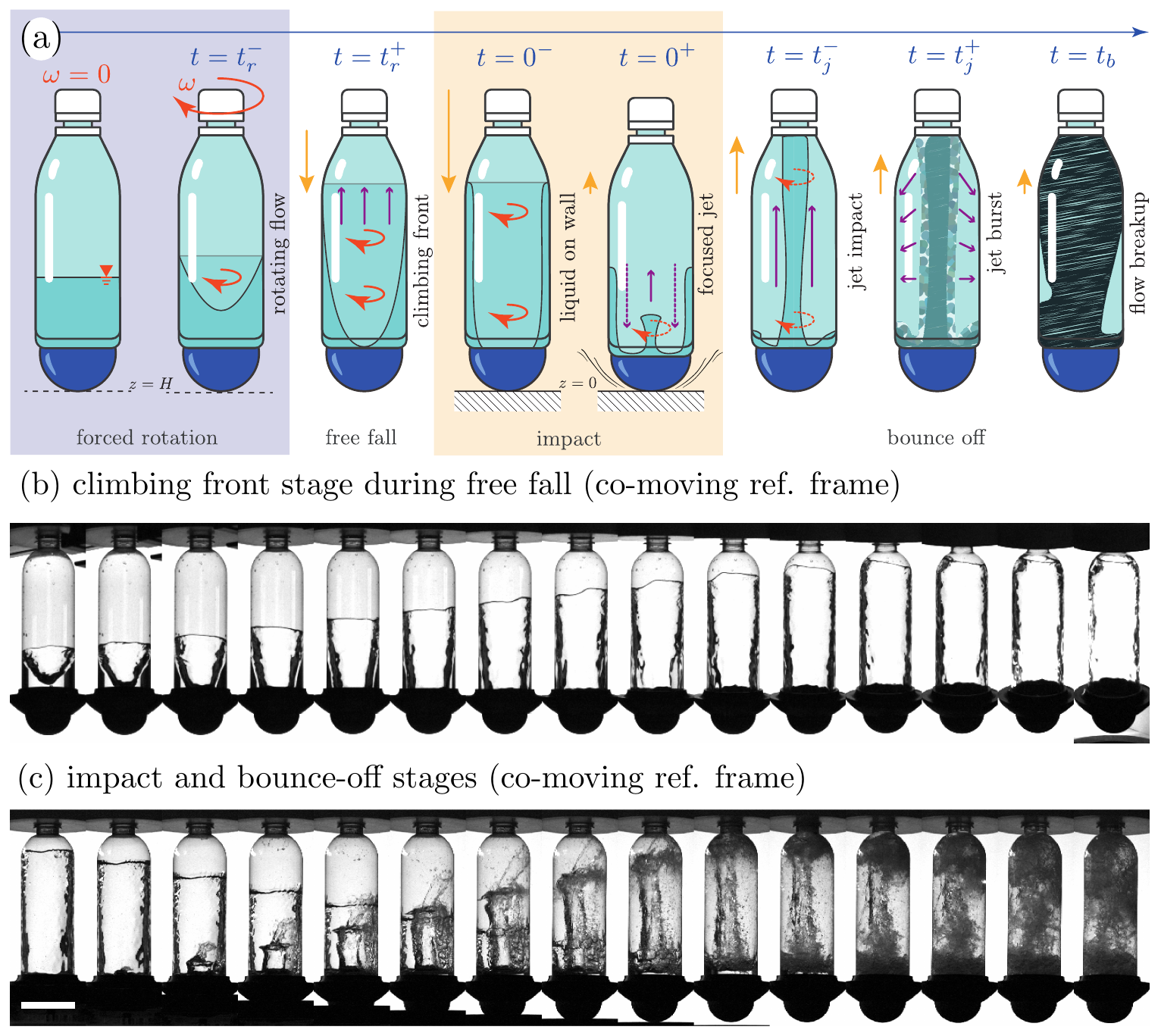}
\par\end{centering}
\caption{Fluid dynamical processes reducing bottle bounce. (a) Four main stages summarize the motion: forced rotation; free fall; impact and bounce-off. Experimental examples are presented in (b) and (c) as image sequences equally spaced in time and shown in the bottle reference frame. (b) Climbing front during free fall from $t_r^-$ to almost $t = 0^-$, spanning 375 ms. (c) Impact and bounce-off in a single sequence going from $t = 0^+$ to $t_b$ in 37 ms. The white bar in panel (c) indicates the bottle diameter of 60 mm.  \label{Fig:StagesScheme}}
\end{figure}
(\emph{iv}) Later during the bounce off ($t=t_j^-$), the head of the central jet hits the top of the container, and a fast turbulent destabilizing front propagates downward along the jet, triggering a fragmentation of the column ($t=t_j^+$). (\emph{v}) Shortly after ($t=t_b$), the breakup has propagated to the entire flow, leaving large regions of drops and blobs full of bubbles that fully block the transmitted light. Note that the experimental follow-up of the  dynamics of this multiphasic stage is overly challenging beyond this point.}
%%. This final stage looks similar to the dynamics after the second bounce in \cite{KillianEtAl_2012}. It should be noted that water and jet dynamics require free space in the container to develop. Therefore, a dependence on the filling volume $\phi$ is also to be expected.}

The rich series of events the rotating liquid undergoes inside the container undoubtedly connects with the reported bounce weakening. A thumb rule is that the central jet carries momentum away. But to what extent? How does rotation itself specifically come into play? What about volume fraction? Does container geometry matter? A precise answer to these questions requires coupling the Navier-Stokes equation for the liquid and the Navier-Cauchy's for the elastic container as an impulsive force acts on the system. Even numerical simulations of this fluid-structure interaction problem will be overly demanding. For these reasons, we instead provide a collision model that successfully captures the key overall outcome of the impacts under study based on physical principles.

\paragraph{} 
To analyze the bottle dynamics and model the impact, we first write the equation of motion in dimensionless form for its vertical position $z(t)$ by considering all the relevant forces
\begin{equation}
\label{Eq:01}
    \ddot{z}=\Pi_ef_e(z)-\Pi_{d}f_{d}(z,\,\dot{z})-\Pi_lf_l(z,\,\dot{z},\,t),
\end{equation}
where $\Pi_e$, measures the nonlinear elastic restitution of the bottom, and $\Pi_d$, a dissipation coefficient that considers energy losses due to contact and deformation. The force between the half sphere and the rigid bottle is given by a contact law $f_{e}\left(z\right)=\left|z\right|^{2}\Theta\left(-z\right)$, where $\Theta$ is the Heaviside function, which properly imposes the contact restraint. The viscoelastic dissipation during the compression of the half sphere is modeled by a nonlinear force $f_{d}(z,\,\dot{z})=\dot{z}\vert z\vert$  \cite{Falcon1998}, which relates to inelastic bounce, i.e. $e<1$, occurring even when the container is empty.  Finally, gravity has been neglected as its effect is markedly weak during the short time of contact. Formulas and justifications of the terms can be found in the Supplemental Material  \cite{SuppMat}.
%%$\Pi_g\equiv gh/v_{-}^2$ represents the gravity pull;  
The last term in Eq.~\eqref{Eq:01} is the key one as it accounts for the fluid forces. $\Pi_l\equiv\rho V_b/m_b$ is an effective mass ratio that compares the mass distribution between liquid and non-liquid parts. 
The analysis of impact dynamics can be simplified by considering the following hypothesis: Just before the impact, the fluid of density $\rho$ rotates at angular frequency $\omega$, and is uniformly distributed on the bottle wall, i.e. in a cylindrical shell with inner and outer radii $r_1$ and $r_2$, and height equal to the full bottle span. This picture is well supported by the experimental evidence of Fig.~\ref{Fig:StagesScheme}(b,c).%, as the fast climbing front always reaches the top before the impact, even at low $\omega$. 

To find an expression for $f_l$, i.e. the fluid force exerted on the bottle, we wrote the conservation laws for the mass and angular momentum assuming an elastic interaction between the bottle and the finite amount of available fluid. During impact, the infinitesimal rotating fluid parcels form a rotating focused jet as they collide with the bottom. The full details can be found in the Supplemental Material  \cite{SuppMat}. Considering that the work rate done on the fluid equals the rate of change of its kinetic energy, we obtain
\begin{widetext}
\begin{equation}        f_{l}\left(z,\,\dot{z},\,t\right)=\phi\cdot\left[\sqrt{\left(\frac{\xi}{\phi}\right)^{2}\left(1-\phi\right)+\left(\left[1+\dot{z}\right]^{2}+\frac{\xi}{2}\right)^{2}}-\left(\frac{\xi}{\phi}\right)+\left(\left[1+\dot{z}\right]^{2}+\frac{\xi}{2}\right)\right]\cdot \Theta\left(1-z-t\right),\label{eq:FL}
\end{equation}
\end{widetext}
where $\xi$ is the ratio between the rotational and translational kinetic energies before impact, and the Heaviside function $\Theta$ accounts for the finite amount of fluid available on the walls. 
Equation~\eqref{eq:FL} shows the explicit dependence of the fluid force on the initial conditions, the filling volume fraction, and the ongoing speed of the container as it hits the target.

\begin{figure}[b]
\begin{centering}
\includegraphics[width=0.95\columnwidth]{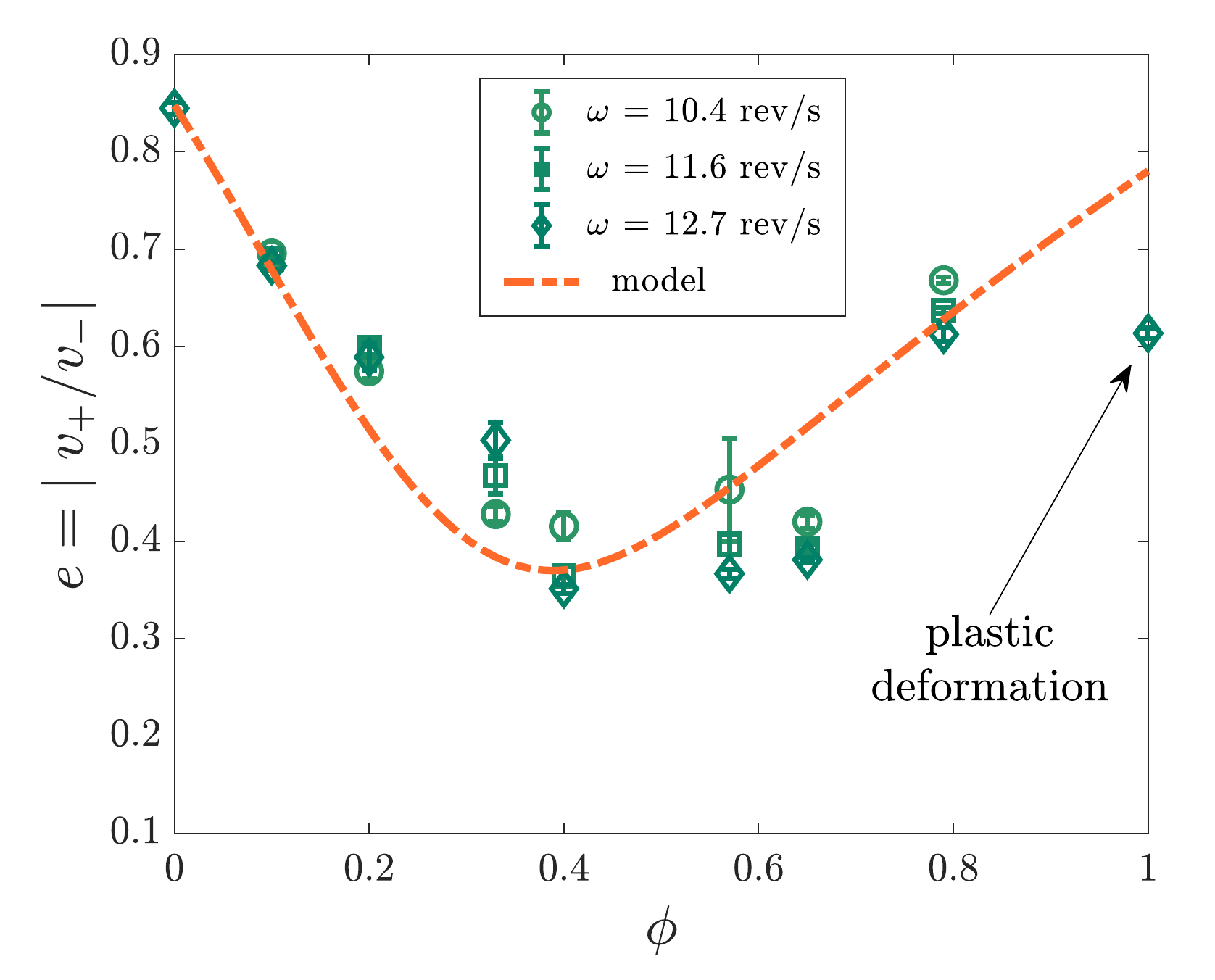}
\par\end{centering}
\caption{Comparison between experimental data and the model. We used $\Pi_l=0.982$, $\Pi_{\Omega}=0.525$, $\Pi_e=27$, and $\Pi_d=0.0583$ for the model, and experimental data for the largest $\omega$ explored, where the modelling hypotheses are fulfilled. The last data point drop is due to container plastic deformations occurring during impact. \label{Fig:Model}}
\end{figure}

%% Dataset 2: L = 0.982, Omega = 0.525, E = 27.0, Delta = 0.0583, sigma = 0.

% Dataset 4:  L = 1.53, Omega = 1.08, E = 27.0, Delta = 0.0583, sigma = 0.

% Inset restitution: L = 0.982, Omega = 0.8, E = 27.0, phi = 0.477, Delta = 0.0583, sigma = 0.

We numerically solved \eqref{Eq:01} for different values of $\phi$ and $\omega$ and obtained $z$-vs.-$t$ curves similar to those shown in Fig.~\ref{Fig:SummaryAndSetup}(c). We also computed the restitution coefficient $e$ from the values of $\dot{z}$ after the bounce. The solid curve in Fig.~\ref{Fig:Model} shows the outcome from the numerical analysis of our model, where $\Pi_d$ and $\Pi_e$ are fitted parameters. Our model captures the overall qualitative and quantitative features of the experimental data, including the emergence of an optimal value for bounce reduction.

\paragraph{}

To summarize, we show that fluid dynamics can significantly reduce the bounce of a partially filled container. We demonstrated this by performing a simple set of experiments: partially filled cylindrical containers set into rotation at frequency $\omega$ and at a given filling volume fraction $\phi$ were released from a fixed height onto a solid target. We study the container bounce via the restitution coefficient $e$ and found large systematic reductions of it. Moreover, we identify optimal values of $\phi$ and $\omega$ for bounce minimization.

The key to understanding the phenomenon is the momentum transfer due to the redistribution of water during impact. After the release of the container, the water set into rotation climbs the walls and redistributes into a cylindrical shell. When the impact occurs, water rapidly focuses into a central jet and gains upward momentum. This transfer generates the decisive stomping force on the container responsible for the great reduction of the bounce. We put this mechanism under test into a collision model, which reproduces the main features of bounce reduction.

Our approach focused in presenting the experimental results of the bounce reduction justifies a framework based on general-physics principles for characterization and modelling. However, rich and complex fluid-dynamic processes arise before, during, and after the impact (see Fig. \ref{Fig:StagesScheme} and videos in \cite{SuppMat}), that deserve further consideration and analysis.  For instance, impact can produce very fast and thick impulsive jets, that carry controllable linear and angular momentum. Also, the fluid breakup is produced not only at jet impact but also during a high-shear stage, when the thick jet is travelling upward.  
Indeed, our experiment has been shown to be remarkably efficient to induce fast flow disintegration of large volumes of liquids in a closed container. 

Although we used a well-controlled experiment, the phenomenon we observed is so robust that it can be readily demonstrated at home by swirling and dropping a partially filled bottle, which we encourage the reader to try. 

%{In order to discuss the dissipation mechanisms, it is useful to consider the works \cite{KigerDuncan_2012,KooijEtAl_2018,Villermaux_2007,EggersVillermaux_2008}.}

%In conclusion, we have evidenced ...

\begin{acknowledgments} 
We acknowledge Tomás Cerda, Francisco Olea and Enriko Granadoz for their help during the early stages of the experiment, Francisco Hauser for technical support, and Enrique Cerda and Eduardo Monsalve for thoughtful discussions. This work was funded by the Agencia Nacional de Investigación y Desarrollo (ANID) through the Fondecyt Grants 11200464 (G.C.), 11190900 (V.S.), 3200499 (J.F.M.), 1221103 (L.G.) and 11191106 (P.G.).
\end{acknowledgments}

\bibliographystyle{apsrev4-1.bst}
%\bibliography{BottleBiblio}
%

\end{document}